\documentstyle [12pt,epsfig]{article}

\begin{document}

\vspace*{2.0cm}

\LARGE

\noindent {\bf Comparison of Two Distinct BBN Codes}

\normalsize

\vspace{2.0cm}

\noindent \hspace{0.8in} {\bf Juan F. Lara}

\vspace{0.4cm}

\small

\noindent \hspace{0.8in} Department of Physics and Center for Relativity,

\noindent \hspace{0.8in} The University of Texas at Austin, TX 78712-1081

\normalsize

\vspace{0.4cm}

\small

\noindent \hspace{0.8in} 
\parbox{5.3in}{{\bf Abstract.}  This paper compares the results of two SBBN
               codes developped independently by different teams of physicists.
               These two codes have significant differences that lead to a 
               discrepency between their final mass fractions of $ ^{4}$He of 
               0.003.  This paper shows that the mass fractions of each code 
               had different orders of convergence, and how the number of 
               timesteps affects the accuracy of the mass fractions.  At the 
               end, the paper shows how to modify both codes so that their
               $ ^{4}$He mass fractions agree to around 0.0001.}

\normalsize

\section*{\normalsize {\bf 1.  Introduction} }

In 1993 I started working on a Big Bang Nucleosynthesis code once used by
UT-Austin Professors Tony Rothman and Richard Matzner ( hereafter known as the
Texas code ) for some papers between 1982 and 1984.\footnote{The papers
dealt with anisotropic models, but the code used for them had the option to
turn the anisotropies off.  Also I expanded this code's original number of
9 isotopes to the other code's 68.}  Two years later Mr. David Thomas
e-mailed me a different BBN code ( hereafter known as the Thomas et al code )
which he used with Profs. David Schramm, Keith Olive, and Mr. Brian Fields for
a series of papers starting in 1993.\footnote{These papers lead up to a 1994
paper on an inhomogenous model, which could be compared to the standard model
of the Thomas et al code.}  But these two different codes calculated different
final mass fractions of their isotopes, as shown in Table (1).  In particular,
the values of $X_{ ^{4}He}$ disagreed by about 0.003.  So I went about 
determining the differences between the codes that could explain this
discrepency.  

A Big Bang Nucleosythesis code starts with free neutrons and protons plus 
photons, electrons and the three neutrino species, all at a high value of 
the electron/photon temperature $T$.  Then in timesteps of value $\Delta t$ the
code calculates the current abundances $Y(i)$ of each isotope as determined
by the reactions between the isotopes and other particles, as well as the 
temperatures and energy densities of the non-baryonic particles, which affect
the reaction rates.  The Thomas et al code featured some details like the
Coulomb factor correction to the neutron-proton conversion rates that I had to
 add to the Texas code.  Also the codes had several differences in calculating
various thermal quantities.\footnote{Kawano ( 1992 ) describes the details of a
code similar to the Thomas et al code.}  The Thomas et al code would for 
instance calculate $T$ via the Runge-Kutta method and then plug in $T$ to 
equations for energy  densities and pressures.  The Texas code would use R-K to
calculate density $\rho_{e + \gamma}$ of the electrons and photons and then 
determine $T$ from there.  I wondered which differences caused the 
disagreements between the codes.  This paper details two differences in the way
each code goes about its evolution that made the accuracy of each code's
results questionable.

\small

\begin{table}[t]

\begin{center}

\begin{tabular}[t]{c|l|l}
  isotope  & \multicolumn{1}{c}{Thomas et al $X$} 
                                   & \multicolumn{1}{c}{Texas $X$} \\ \hline
     d     & $0.1203806246 \times 10^{-3}$  & $0.1239148256 \times 10^{-3}$ \\
 $ ^{3}$H  & $0.5791818972 \times 10^{-6}$  & $0.5993673829 \times 10^{-6}$ \\
 $ ^{3}$He & $0.3616593549 \times 10^{-4}$  & $0.364445127 \times 10^{-4}$ \\
 $ ^{4}$He & $0.2369001572$                 & $0.2397015878$ \\    
 $ ^{7}$Li & $0.3349991579 \times 10^{-9}$  & $0.3496296345 \times 10^{-9}$ \\
 $ ^{7}$Be & $0.286454318 \times 10^{-9}$   & $0.2679050102 \times 10^{-9}$ \\
 $ ^{9}$Be & $0.1036417579 \times 10^{-16}$ & $0.1126556133 \times 10^{-16}$ \\
 $ ^{10}$B & $0.8927007279 \times 10^{-18}$ & $0.9643350569 \times 10^{-18}$ \\
 $ ^{11}$B & $0.9314884547 \times 10^{-16}$ & $0.91442805 \times 10^{-16}$ \\
 $ ^{12}$C & $0.2898262833 \times 10^{-13}$ & $0.3048347851 \times 10^{-13}$ \\
 $ ^{14}$N & $0.1275229996 \times 10^{-12}$ & $0.1326174221 \times 10^{-12}$ \\
 $ ^{16}$O & $0.7290963822 \times 10^{-17}$ & $0.7730848001 \times 10^{-17}$ \\
$ ^{21}$Ne & $0.6698206755 \times 10^{-21}$ & $0.5517307027 \times 10^{-23}$ \\
\end{tabular}

\vspace*{0.2in}

   {\bf Table (1)}: Mass Fractions $X$ from each code.  The results are for 
$\eta_{10} = $ 3.0, but the codes have the same discrepency between the values
of $X_{ ^{4}He}$ for all interesting values of $\eta_{10}$.

\end{center}

\end{table}

\section*{\normalsize {\bf 2.  Convergence of the codes} }

\normalsize

At each timestep $n$ the BBN codes use the Second-Order Runge-Kutta method to
calculate the isotope abundances $Y_{n+1}(i)$.:

\begin{eqnarray}
   Y_{n+1}(i) & = & Y_{n}(i) + \frac{1}{2} \left [ \frac{dY(i)}{dt}
                    \left ( t, Y_{n} \right ) + \frac{dY(i)}{dt}
                    \left ( t + \Delta t, Y_{n+1} \right ) \right ] \Delta t
\end{eqnarray}

\noindent Wagoner ( 1969, p. 253 ) lists the complicated equation for 
$\dot{Y}(i)$, which depends on the other abundances and on the reaction rates
that destroy ( $[ij]$ ) and create ( $[kl]$ ) isotope $i$.  But at the 
highest temperatures the total destruction rate of each isotope is nearly equal
to its creation rate.  So the codes calculate $\dot{Y}(i)$ using an equation
linearized in terms of $Y_{n+1}(i)$

\begin{eqnarray*}
   \frac{dY(i)}{dt} ( t, Y_{n} ) & = & 
                 A_{ij} ( Y_{n}, [ij](T_{n}), [kl](T_{n}) ) \times Y_{n+1}(j) 
\end{eqnarray*}

\noindent where the matrix $A_{ij}$ is written out in Wagoner (1969, p. 294 ).
The codes pair off the right hand side of this equation with an implicit 
expression of $\dot{Y}(i)$ leading to a matrix equation for $Y_{n+1}(i)$.

\pagebreak

\begin{eqnarray}
   Y_{n}(i) & = & Y_{n+1}(i) - \frac{dY(i)}{dt}[t+\Delta t_{n}, Y_{n+1}(i) ]
                  \Delta t_{n} + \cdots \nonumber \\
   \frac{dY(i)}{dt}[t+\Delta t_{n}, Y_{n+1}(i) ] 
            & = & \frac{Y_{n+1}(i) - Y_{n}(i)}{\Delta t_{n}} \\
   Y_{n}(i) & = & Q_{ij} ( Y_{n}, [ij](T_{n}), [kl](T_{n}) ) \times
                  Y_{n+1}(j) 
\end{eqnarray} 

\noindent ( $Q_{ij} = 1_{ij} - A_{ij} \Delta t_{n}$ ) The codes solve for
$Y_{n+1}(i)$ and plug it back in to get $\dot{Y}(i)$.

This R-K method should produce mass fractions that have second-order 
convergenge as the timestep values get smaller.  I tested the convergences of
mass fraction $X_{ ^{4}He}$ the Thomas et al code and the Texas code by 
starting with an array of timestep values $\Delta t_{n}$ that naturally arise
from a run, and then dividing those values by two, four, and so on.  The 
Thomas el al code turned out to have the expected convergence:

\begin{eqnarray*}
   X_{ ^{4}He} & = & 0.2392602277 - ( 0.2474541681 \times 10^{-2} ) d^{2}
\end{eqnarray*} 

\noindent where $d$ equals one over the number of divisions of each timestep.
But the Texas code had only first order convergence, and so needed to be 
fixed.

\begin{eqnarray*}
   X_{ ^{4}He} & = & 0.2394509836 + ( 0.3832102957 \times 10^{-3} ) d
\end{eqnarray*} 

I traced each code to find out exactly how in step \# $n$ each went about 
calculating $Y_{n+1}(i)$.  In the first Runge-Kutta step the Thomas et al code
used the previous $Y_{n-1}(i)$ on the left hand side of matrix equation (3), 
and $Y_{n} (i)$ in matrix $Q_{ij}$.  

\begin{eqnarray*}
       Y_{n-1}(i) & = & Q_{ij} ( Y_{n}, [ij](T_{n}), [kl](T_{n}) ) \times 
                        Y_{n\alpha}(j) \\
   \frac{dY(i)}{dt} ( t, Y_{n}(i) ) 
                  & = & \frac{Y_{n\alpha}(i) - Y_{n-1}(i)} {\Delta t_{n-1}}
\end{eqnarray*}

\noindent I'm calling the equation solution $Y_{n\alpha}(i)$ for the first
step ( and then $Y_{n\beta}(i)$ for the second step ).  Then this code used
an expression for $\dot{Y}(i)$ that clearly matched up with Equation (2) for
$\dot{Y}(i) [ t, Y_{n}(i) ]$, with $Y_{n-1}(i)$ in the numerator and 
$\Delta t_{n-1}$ in the denominator.  So $Y_{n\alpha}(i)$ appears to be an
approximation of $Y_{n}$.  The Thomas et al code then uses 
$\dot{Y}(i) [ t, Y_{n}(i) ]$ and a new timestep $\Delta t_{n}$ to get the 
value $\tilde{Y}_{n}(i)$

\begin{eqnarray*}
 \tilde{Y}_{n}(i) & = & Y_{n}(i) + 
                        \frac{Y_{n\alpha}(i) - Y_{n-1}(i)}{\Delta t_{n-1}}
                        \Delta t_{n} 
\end{eqnarray*}
\noindent an approximation of $Y_{n+1}(i)$ to be used in the matrix of the
second Runge-Kutta step.

\begin{eqnarray}
     Y_{n}(i) & = & Q_{ij} \left ( \tilde{Y}_{n}, [ij](\tilde{T}_{n}), 
                    [kl](\tilde{T}_{n}) \right ) \times Y_{n\beta}(j) 
                    \nonumber \\
   \frac{dY(i)}{dt} ( t+\Delta t_{n}, Y_{n+1}(i) ) 
              & = & \frac{Y_{n\beta}(i) - Y_{n}(i)}{\Delta t_{n}} \nonumber \\
   Y_{n+1}(i) & = & Y_{n}(i) + \frac{1}{2} \left [ 
                    \frac{Y_{n\alpha}(i) - Y_{n-1}(i)}{\Delta t_{n-1}} +
                    \frac{Y_{n\beta}(i) - Y_{n}(i)}{\Delta t_{n}} \right ]
                    \Delta t_{n} 
\end{eqnarray}

\noindent And this codes used $Y_{n}(i)$ in its matrix equation.  So one could
say that $Y_{n\beta}$ was an approximation of $Y_{n+1}(i)$, and this second
time derivative corresponded to $\dot{Y}(i)[ t+\Delta t_{n}, Y_{n+1}(i) ]$
according to Equation (2).  Plugging them into Equation (1) we get Equation (4)
for our actual value of $Y_{n+1}(i)$.  Note the $\Delta t_{n}$ as a factor and
the $\Delta t_{n-1}$ in one of the denominators.  

The first R-K step of the Texas code, though, has $Y_{n}(i)$ on the left-hand
side, and $\Delta t_{n}$ in the denominator for $\dot{Y}(i)$.:

\begin{eqnarray*}
           Y_{n}(i) & =       & Q_{ij} ( Y_{n}, [ij](T_{n}), [kl](T_{n}) ) 
                                \times Y_{n\alpha}(j) \\
   \frac{dY(i)}{dt} \left ( t, Y_{n} \right ) 
                    & =     & \frac{Y_{n\alpha}(i) - Y_{n}(i)}{\Delta t_{n}} \\
\end{eqnarray*}

\noindent That would imply that this code's $Y_{n\alpha} (i)$ is an 
approximation of $Y_{n+1} (i)$, and that $\dot{Y}(i) ( t, Y_{n} )$ has been
determined explicitly instead of implicitly.  The second R-K step, 

\begin{eqnarray*}
     Y_{n}(i) & = & Q_{ij} ( \tilde{Y}_{n}, [ij](T_{n+1}), [kl](T_{n+1}) ) 
                    \times Y_{n\beta}(j) \\
  \frac{dY(i)}{dt} ( t+\Delta t_{n}, \tilde{Y}_{n} ) 
              & = & \frac{Y_{n\beta}(i) - Y_{n}(i)}{\Delta t_{n}} \\
\end{eqnarray*}

\noindent implies that $Y_{n\beta}$ is also an approximation of $Y_{n+1}$,
and when the Texas code plugs in our expressions for the $\dot{Y}(i)$'s.:

\begin{eqnarray}
   Y_{n+1}(i) & = & Y_{n}(i) + \frac{1}{2} \left [ 
                    \frac{Y_{n\alpha}(i) - Y_{n}(i)}{\Delta t_{n}} +
                    \frac{Y_{n\beta}(i) - Y_{n}(i)}{\Delta t_{n}}
                    \right ] \Delta t_{n} \nonumber \\
   Y_{n+1}(i) & = & \frac{1}{2} ( Y_{n\alpha}(i) + Y_{n\beta}(i) )
\end{eqnarray}

\noindent the $\Delta t_{n}$ factor cancels out the $\Delta t_{n}$'s in the
denominators.  So the final $Y_{n+1} (i)$ is the average of the two first-order
Euler approximations $Y_{n\alpha}$ and $Y_{n\beta}$, calculated from solving
the matrix equations alone.  

So I modified the Texas code to follow the Thomas et al code's more consistent
scheme of calculating $Y_{min}$.  The Texas code's $X_{ ^{4}He}$ then acquired
the second-order convergence it should've had.:

\begin{eqnarray*}
   X_{ ^{4}He} & = & 0.2394508633 - ( 0.1509902093 \times 10^{-3} ) d^{2}
\end{eqnarray*} 

\noindent For $d = $ 1, $X_{ ^{4}He}$ fell from 0.23970 to 0.23929.  Still
far from the other code's 0.23690.  But $X_{ ^{4}He}$ in the Texas code 
converged to the same value as it did with first-order.

\section*{\normalsize {\bf 3.  The value of $Y_{min}$} }

$Y(i)$ of the Texas code could go all the way down to 0.0.  But the Thomas et
al code put a lower limit $Y_{min} = 10^{-25}$.  That code used $Y_{min}$ in
its equation for $\Delta t_{n}$

\pagebreak

\begin{figure}[t]
   \leftline{\epsfxsize=3.0in\epsffile{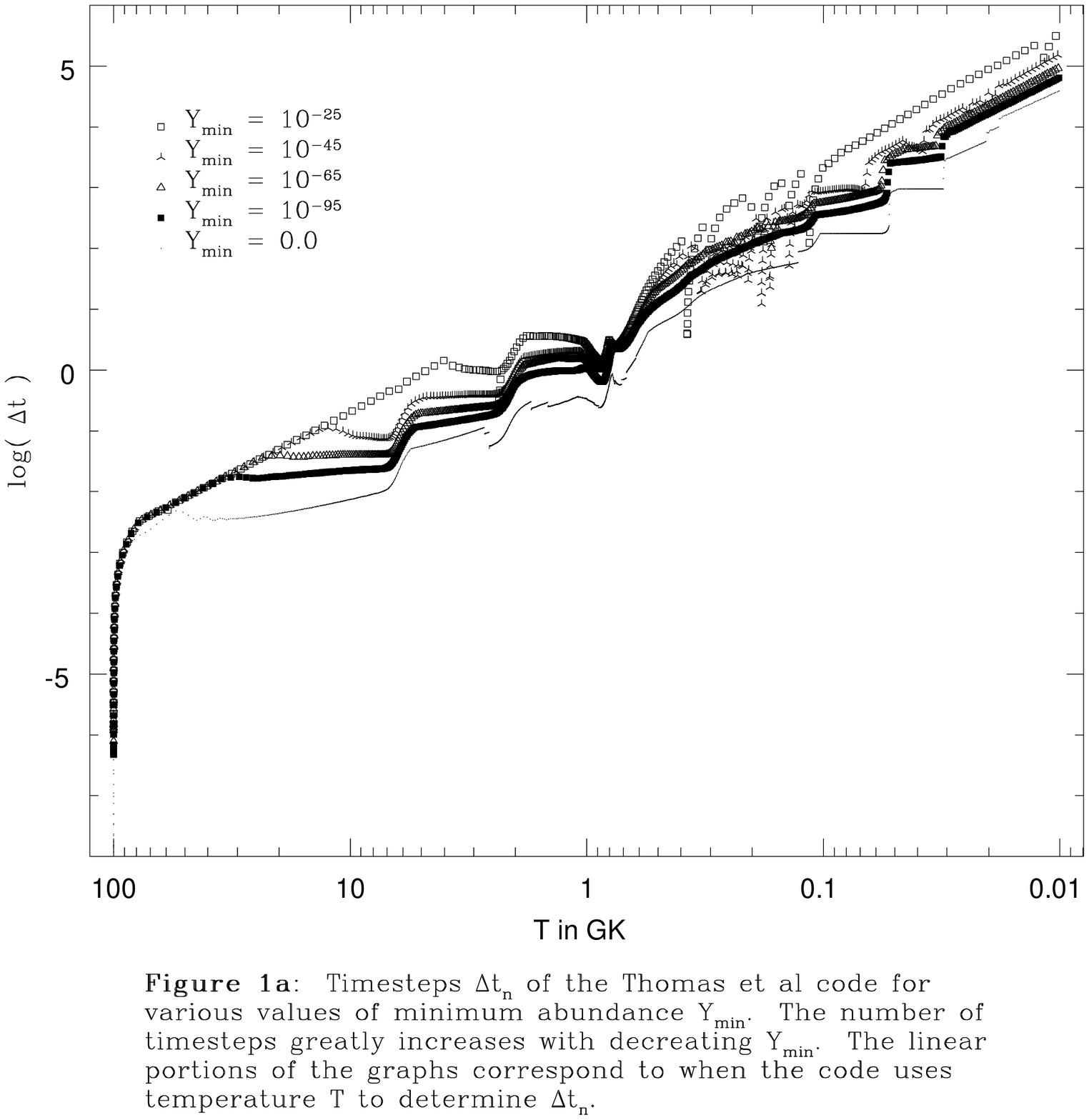}
             \epsfxsize=3.0in\epsffile{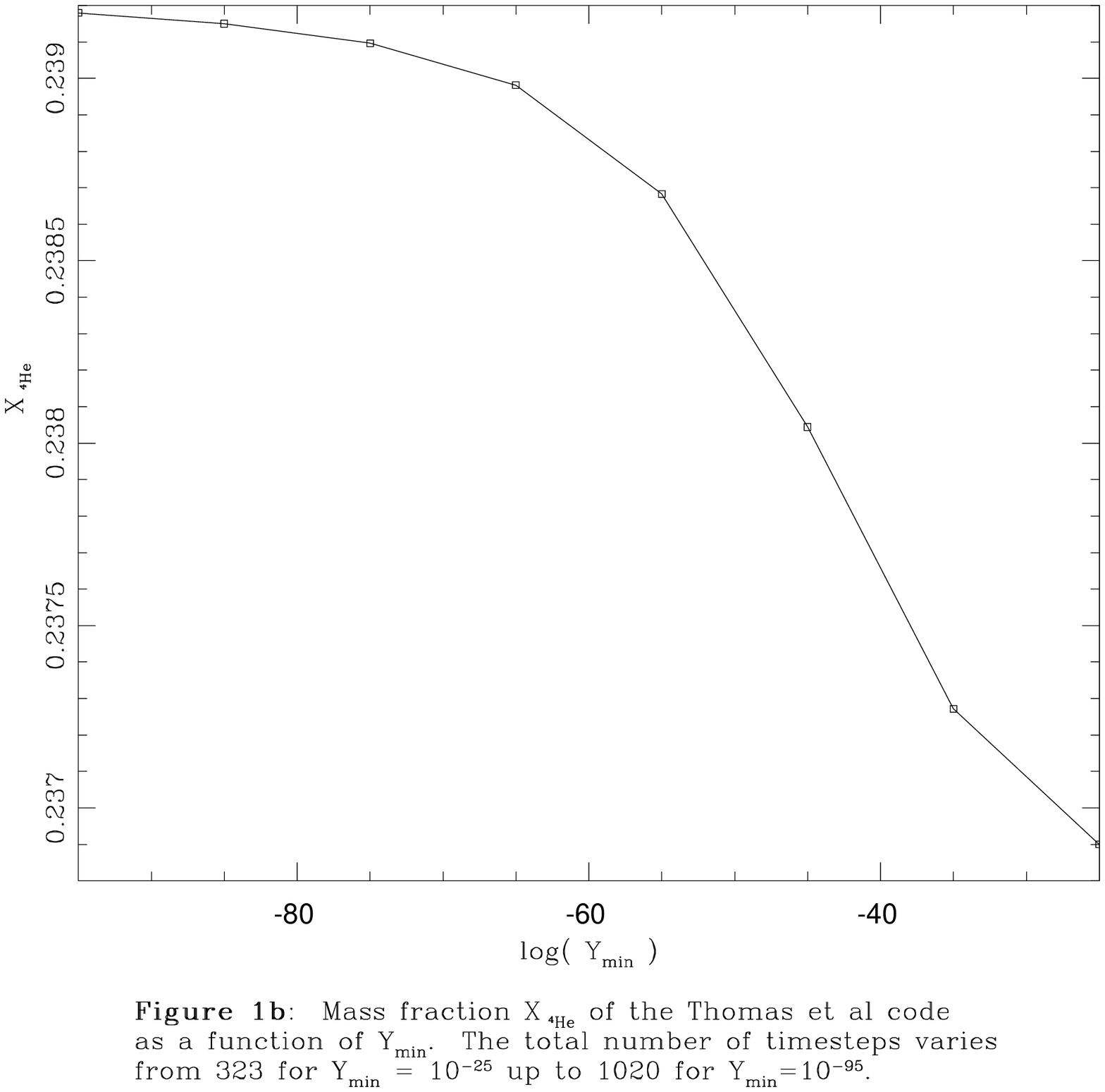}}
\end{figure}

\begin{eqnarray*}
    \Delta t_{n} & = & dk \min \left [ \frac{Y_{n}(i)}{\dot{Y}(i)} dk_{fac},
                                       \frac{T_{n}}{\frac{dT}{dt}} \right ] \\
        dk_{fac} & = & \left ( 1 + \left ( -\frac{\log Y_{n}(i)}{dt_{fac}} 
                       \right )^{2} \right ) \\
        dt_{fac} & = & -\frac{\log Y_{min}}{\frac{dk_{max}}{dk} - 1} 
\end{eqnarray*}   

\noindent The Thomas et al code compared the ratio of $T$ to $\dot{T}$ to 
the $Y(i)/ \dot{Y}(i)$ ratios to pick the smallest ratio times $dk = $ 0.1 as
the value of $\Delta t_{n}$, so that the code could evolve stably. But to the
$Y(i)/ \dot{Y}(i)$ ratio the code also put a factor $dk_{fac}$, designed to
prevent the code from picking a $Y(i)$ whose value was close to $Y_{min}$.  
But this factor lead to larger values of $\Delta t_{n}$ than in the Texas code,
and hence much fewer timesteps.  At $\eta_{10} = 3.0$, the 
\begin{figure}[h]
   \leftline{\epsfxsize=3.0in\epsffile{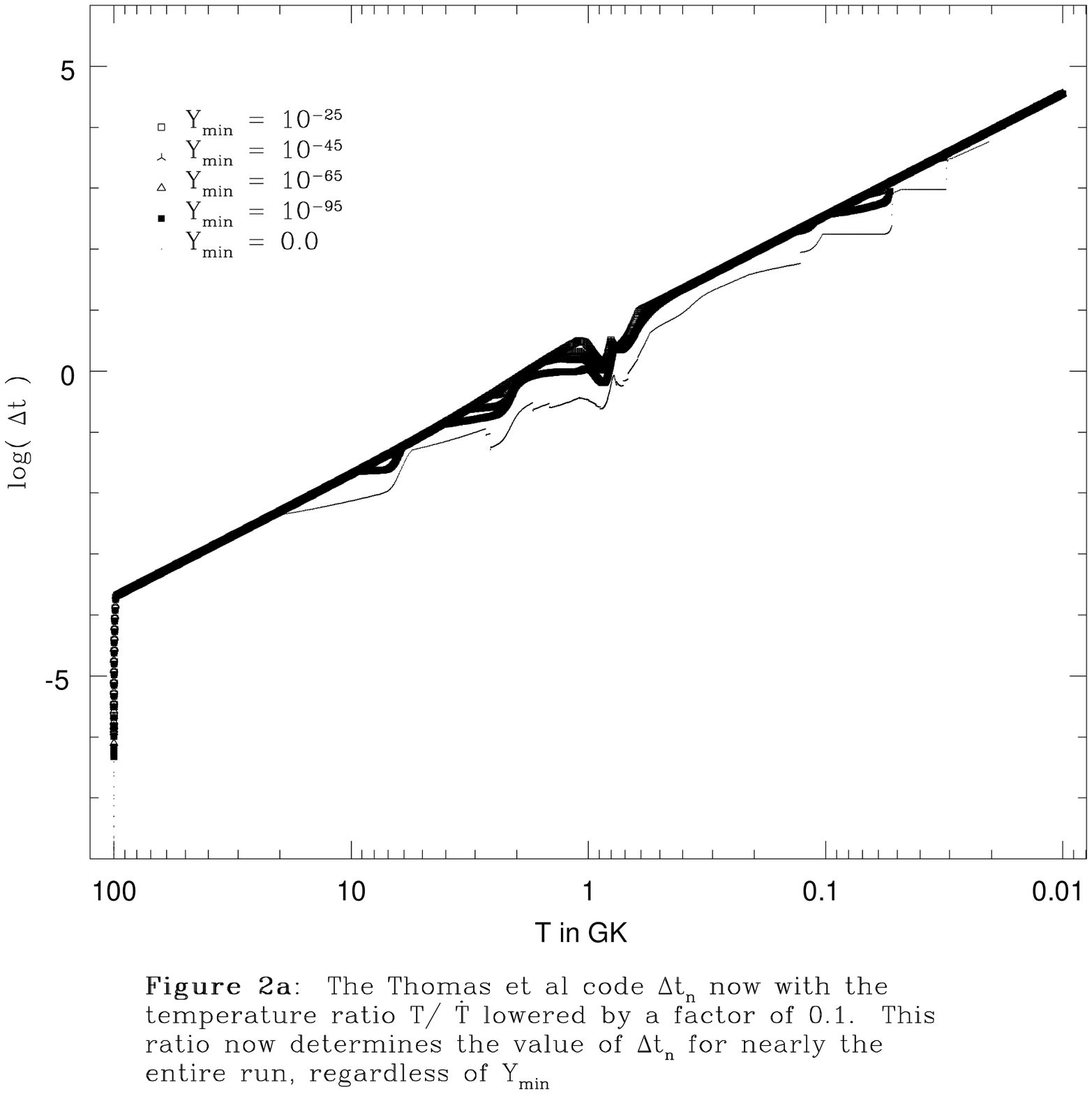}
             \epsfxsize=3.0in\epsffile{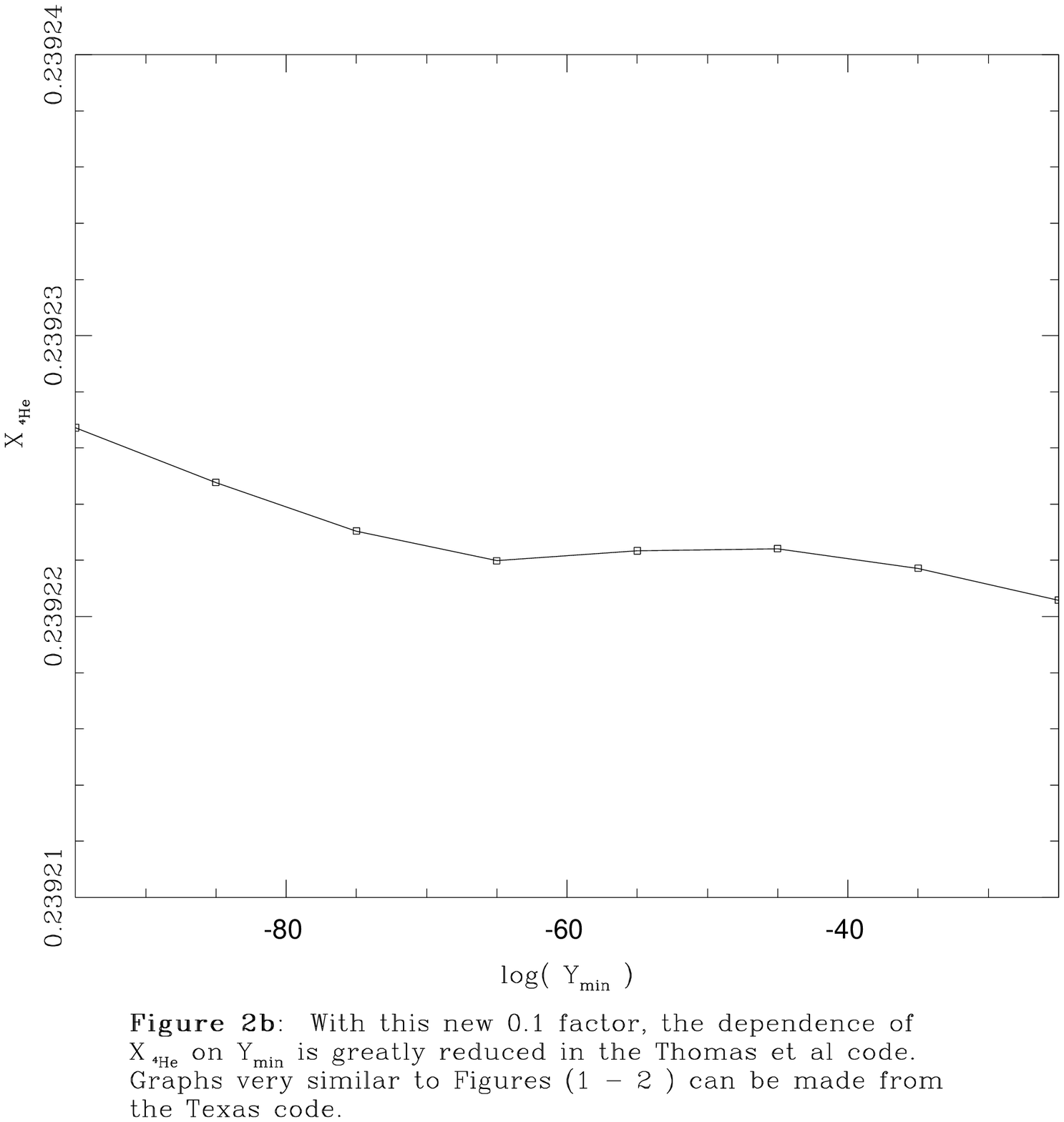}}
\end{figure}

\pagebreak

\noindent Thomas et al code 
would run in only 323 timesteps, whereas the Texas code needed 2772.  Figure 
(1a) graphs $\Delta t_{n}$ for various values of $Y_{min}$.  Varying $Y_{min}$
from $10^{-25}$ to $10^{-95}$ would vary the number of timesteps from around
300 to 1000.  And $X_{ ^{4}He}$ would vary from 0.257 to 0.259, as shown in
Figure (1b).

The linear sections in Figure (1a) corresponded to when the code chose 
$T/ \dot{T}$ to determine $\Delta t_{n}$.  So I got the idea of putting on
$T/ \dot{T}$ a second factor 0.1.  Figure (2a) shows that this 0.1 factor
lowered the $T/ \dot{T}$ ratio to the point that the code would choose that
ratio nearly all of the time.  So in Figure (2a) the number of timesteps varied
only from 1000 to 1300.  And Figure (2b) shows that the 0.1 factor 
eliminated the influence of $Y_{min}$ on $X_{ ^{4}He}$.  So the number of 
timesteps instead of $Y_{min}$ itself really determined the accuracy of 
$X_{ ^{4}He}$.

The Texas code uses this equation to determine the timestep.

\begin{eqnarray*}
   \Delta t_{n+1} & = & dk \min \left [ \frac{Y_{n+1}(i)}{\frac{dY(i)}{dt}},
                        \frac{( ln R )_{n+1}}{\frac{\dot{R}}{R}},
           \frac{( \rho_{e + \gamma} )_{n+1}}{\dot{\rho}_{e + \gamma}} \right ]
\end{eqnarray*}

\noindent  No $dk_{fac}$ factor.  So the code could run with $Y_{min}$ set at
zero.  I modified the Texas code to have a $dk_{fac}$ for a non-zero $Y_{min}$.
I also wanted to put in a 0.1 factor, but instead of $T$ the Texas code used
$\rho_{e+\gamma}$, which depended on $T^{4}$ for most of a run.  So I put in
a 0.4 factor instead.  And indeed the Texas code exhibited the same behavior
as the Thomas et al code with these new conditions, though $X_{ ^{4}He}$ tended
to vary more here than in the other code.

I'd still prefer to not have a non-zero $Y_{min}$ at all.  But the Texas code's
old way of calculating $\Delta t_{n}$ would have that large number of steps in
that case.   So for $Y_{min} = $ 0.0 I put in both codes the following 
$dt_{fac}$ factor.:

\begin{eqnarray*}
         dk_{fac} & = & \left ( 1 + \left ( -\frac{\log Y_{n}(i)}{dt_{fac}} 
                        \right )^{2} \right ) \nonumber \\
         dt_{fac} & = & -\frac{\log 10^{-99}}{\frac{dk_{max}}{dk} - 1} 
\end{eqnarray*}   

This factor set the number of timesteps at 1320 for both codes.  A number on
the order of 1000 seemed sufficient to get a value of $X_{ ^{4}He}$ that 
didn't depend very much on the timestep number.

\section*{\normalsize {\bf 4.  Final results} }

Table (2) shows the final mass fractions for each code with the modifications
put in, for the case of $\eta_{10} = $ 3.0.  The codes now have close 
agreement with each other, especially for $ ^{4}$He where the mass fractions
are within 0.0001 of each other.  These results are for $Y_{min} = 0.0$, but I
got similar results for $Y_{min} = 10^{-25}$ as well.

As for the thermal quantities, those differences seemed very confusing.  But I
checked these thermal quantities and determined them to be nearly equal between
the codes after  I put the changes in.  So the number of timesteps and the 
convergences were the most signficant reasons for disagreement between the 
codes.  

\begin{table}[t]

\begin{center}

\small

\begin{tabular}[t] {c|l|l}
   isotope  & \multicolumn{1}{c}{Thomas et al $X$} 
                                   & \multicolumn{1}{c}{Texas $X$} \\ \hline
     d     & $0.1214409373 \times 10^{-3}$  & $0.1222110665 \times 10^{-3}$ \\
 $ ^{3}$H  & $0.5869118485 \times 10^{-6}$  & $0.5905659297 \times 10^{-6}$ \\
 $ ^{3}$He & $0.3615054989 \times 10^{-4}$  & $0.3624197334 \times 10^{-4}$ \\
 $ ^{4}$He & $0.2392346716$                 & $0.2393288161$ \\    
 $ ^{7}$Li & $0.3424567551 \times 10^{-9}$  & $0.3445304655 \times 10^{-9}$ \\
 $ ^{7}$Be & $0.2810372859 \times 10^{-9}$  & $0.277271461 \times 10^{-9}$ \\
 $ ^{9}$Be & $0.1078010758 \times 10^{-16}$ & $0.1091991775 \times 10^{-16}$ \\
 $ ^{10}$B & $0.9337145278 \times 10^{-18}$ & $0.9425546233 \times 10^{-18}$ \\
 $ ^{11}$B & $0.9855540716 \times 10^{-16}$ & $0.9595478082 \times 10^{-16}$ \\
 $ ^{12}$C & $0.3026820198 \times 10^{-13}$ & $0.30138561 \times 10^{-13}$ \\
 $ ^{14}$C & $0.6519717631 \times 10^{-13}$ & $0.6515513573 \times 10^{-13}$ \\
 $ ^{14}$N & $0.1322155965 \times 10^{-12}$ & $0.1320911107 \times 10^{-12}$ \\
 $ ^{16}$O & $0.7673887093 \times 10^{-17}$ & $0.7632076774 \times 10^{-17}$ \\
$ ^{21}$Ne & $0.5388613183 \times 10^{-23}$ & $0.5374132790 \times 10^{-23}$ \\
\end{tabular}

\vspace*{0.2in}

{\bf Table (2)}:  Mass Fractions for Modified Codes  Here, $Y_{min} = $ 0.0
and $\eta_{10} = $ 3.0.  But the codes still agree for the range of 1 to 10
for $\eta_{10}$ and $Y_{min}$ up to $10^{-25}$.

\end{center}

\end{table}

\normalsize

And finally, I took convergence tests of the codes again after all the 
changes had been added.  The Thomas et al code obeyed:

\begin{eqnarray*}
   X_{ ^{4}He} & = & 0.2392673182 - ( 0.3253724834 \times 10^{-4} ) d^{2}
\end{eqnarray*} 

\noindent while the Texas code obeyed.

\begin{eqnarray*}
   X_{ ^{4}He} & = & 0.2394508628 - ( 0.1203603394 \times 10^{-3} ) d^{2}
\end{eqnarray*} 

The codes still converged to the same values that they've converged before.
So the remaining differences between the codes result in these final values
disagreeing by 0.0002.

\section*{\normalsize {\bf References} }

\small

Kawano, L.  Fermilab-Pub-92/04-A ( 1992 )

\vspace*{0.15in}

\noindent
Rothman, T., Matzner, R.  Physical Review Letters {\bf 48}, 1565 (1982).

\vspace*{0.15in}

\noindent
Rothman, T., Matzner, R.  Physical Review D. {\bf 30}, 1649 (1984).

\vspace*{0.15in}

\noindent
Thomas, D., Schramm. D., Olive, K.A., Fields, B.D.  Astrophysical Journal
{\bf 406} 569 ( 1993 )

\vspace*{0.15in}

\noindent
Thomas, D., Schramm. D., Olive, K.A., Mathews, G.J., Meyer, B.S., Fields, B.D.

Astrophysical Journal {\bf 430} 291 ( 1994 )

\vspace*{0.15in}

\noindent
Wagoner, R. V.  Astrophysical Journal Supplement Series {\bf 18}, 247 (1969).

\end{document}